# Walk Entropies in Graphs

Ernesto Estrada,[1,*] José A. de la Peña,[2] and Naomichi Hatano[3]

[1]Department of Mathematics and Statistics, University of Strathclyde, Glasgow G1 1XH, U.K.; [2]CIMAT, Guanajuato, Mexico; [3]Institute of Industrial Science, University of Tokyo, Komaba 4-6-1, Meguro, Tokyo 153-8505, JAPAN

**Abstract**

Entropies based on walks on graphs and on their line-graphs are defined. They are based on the summation over diagonal and off-diagonal elements of the thermal Green's function of a graph also known as the communicability. The walk entropies are strongly related to the walk regularity of graphs and line-graphs. They are not biased by the graph size and have significantly better correlation with the inverse participation ratio of the eigenmodes of the adjacency matrix than other graph entropies. The temperature dependence of the walk entropies is also discussed. In particular, the walk entropy of graphs is shown to be non-monotonic for regular but non-walk-regular graphs in contrast to non-regular graphs.

PACS numbers: 89.70.Cf, 05.30.-d, 02.10.Ox, 03.67.-a

*Corresponding author. E-mail: ernesto.estrada@strath.ac.uk

# 1. Introduction

With recent surge of interest in complex networks in various fields including statistical physics and mathematical physics, many quantities have been proposed to characterize the structural properties of graphs [1, 2]. The study of a graph invariant in one field may also be a result of relevant importance in other areas of physics. This is because graphs are nowadays ubiquitous in many areas of physics such as in problems associated with the Ising, Potts and Hubbard models, in the solution of Feynman integrals in perturbative field theory, in quantum information theory such as quantum error correcting codes (graph states) or arrangements of interacting quantum mechanical particles (spin networks) [3-7], and in many other fields (see [8] and references therein).

Among various graph invariants, a special role has been played by the concept of entropy. Entropy measures for graphs have been used for a long time in different fields [9-12]. Inspired by connections between quantum information and graph theory, Passerini and Severini [10] have defined the von Neumann entropy for graphs, which in general depends on the regularity, the number of connected components, the shortest-path distance and nontrivial symmetries in the graph. This entropy is defined on the basis of the eigenvalues of the discrete Laplacian matrix $L$ of a graph: $S = -\sum_{j=1}^{n} \mu_j \log \mu_j$, where $\mu_j$ is an eigenvalue of $L$. Previously, Estrada and Hatano [11] have defined the Shannon entropy of a network by using a tight-binding Hamiltonian of the form $H = -A$, where $A$ is the adjacency matrix of the graph. That entropy is based on the probability $p_j = \exp(\lambda_j)/Z$ of finding the graph in a state with energy given by $-\lambda_j$, where $\lambda_j$ is an eigenvalue of $A$ and $Z = \sum_{j=1}^{n} \exp(\lambda_j)$.

Here, we define graph entropies based on walks in a graph. Walks in graphs play a fundamental role in the analysis of the structure and dynamical processes in networks [13]. The new graph entropies, namely the walk entropies, account for the amount of uncertainty in selecting a walk that started (and ended) at a given node or edge of the graph. The walk entropies thereby characterize the spread of a walk among the vertices or edges of the graph; in other words, we understand by the walk entropies how much the walk is concentrated, or "localized" in just a few nodes. We show that the behavior of the walk entropies is remarkably different for walk-regular, regular and non-regular graphs. The walk entropies have their maximum for the walk-regular graphs, which include important graphs such as vertex-transitive graphs, distance-regular graphs and strongly regular graphs [14]. Some of these graphs, namely distance-regular and strongly-regular ones, have been studied in the context of quantum information theory with different interesting properties [15-20]. We also analyze the effects of the temperature on the walk entropies and the localization in different types of graphs. We introduce the walk entropy for a graph in Sec. 2 and for the line graph of a graph in Sec. 3. In Sec. 4, we relate the walk entropies to the localization of a walker on the nodes and edges of a graph. Section 5 further argues the temperature effect on the relation between the walk entropies and the localization.

Before proceeding, we summarize a few definitions which are necessary to make this paper self-contained. Let us consider here simple graphs $G = (V, E)$ with $|V| = n$ nodes and $|E| = m$ edges; no multiedges or self-loops are allowed. A walk of length $k$ is a sequence of (not necessarily distinct) nodes $v_0, v_1, \cdots, v_{k-1}, v_k$ such that for each $i = 1, 2 \cdots, k$ there is a link from $v_{i-1}$ to $v_i$. If $v_0 = v_k$, the walk is named a closed walk. The number of walks of length $k$ from node $p$ to node $q$ is given by $(A^k)_{pq}$, where $A$ is the adjacency matrix of the graph. A graph is said to be *regular* if every node has the same degree. The degree of the node $p$, denoted by $k_p$, is the number of edges incident to it. A *walk-regular graph* is a graph for which $(A^k)_{pp} = \omega$ for any $k$ and for all nodes of the graph, where $\omega$ is a certain integer number. In other words, the number of the closed walks is the same for any $p$ and $k$. It is known that a walk-regular graph is also regular.

In order to define graph entropies based on the walks, we consider a random walker which walks from one node to another by using the edges of the graph. This consideration is similar to the ones previously used in Ref. [15] and more recently in several works [16-18]. We identify the negative adjacency matrix as a Hamiltonian of the walker and consider the thermal Green's function of the graph as previously described by Estrada and Hatano [21]

$$G_{pp}(b) = \langle p|e^{-bH}|p\rangle = \langle p|e^{bA}|p\rangle, \tag{1}$$

where $\beta = (k_B T)^{-1}$ is the inverse temperature. The partition function for the graph is then defined by [11, 13]

$$Z(\beta) = tr(e^{\beta A}). \tag{2}$$

Since we can expand the matrix $e^{\beta A}$ in the form

$$(e^{\beta A})_{pq} = \sum_{k=0}^{\infty} \frac{\beta^k (A^k)_{pq}}{k!}, \tag{3}$$

it counts all walks from one node to another in the graph in a way that the walks of length $k$ are penalized by the inverse of the factorial of its length and hence the shorter walks receive more weight than the longer ones.

## 2. Walk Entropies

We start by defining the probability of selecting at random a closed walk that has started (and ended) at the node $i$, among all the closed walks in the graph. That is,

$$p_i(\beta) \stackrel{def}{=} \frac{(e^{\beta A})_{ii}}{Z}. \tag{4}$$

Using the Shannon formula we now define the walk entropy based on the nodes of the graph as follow

$$\begin{aligned} S^V(G,\beta) &\stackrel{def}{=} -\sum_i \frac{(e^{\beta A})_{ii}}{Z} \log_2 \frac{(e^{\beta A})_{ii}}{Z} \\ &= -\sum_i \frac{(e^{\beta A})_{ii}}{Z} \left[ \log_2 (e^{\beta A})_{ii} - \log_2 Z \right] \end{aligned} \tag{5}$$

We calculate the walk entropy for all possible 11,117 connected graphs with 8 nodes. In contrast with the von Neumann entropy [10] and the Shannon entropy previously developed [11], the walk entropy is not correlated with the number of edges in the graphs. For instance, the walk entropy displays a poor correlation coefficient of 0.14, while the von Neumann and Shannon display very high correlation coefficients, 0.84 and $-0.94$, respectively. In other words, the walk entropy is free of a strong dependence with the number of edges in the graph, which may be important for analyzing the structure of much larger networks.

It is straightforward to realize that for $0 < \beta < \infty$ a graph $G$ with $n$ nodes has the maximum entropy $\max_n S^V(G,\beta) = \log_2 n$ if it is walk-regular. We may conjecture that if $S^V(G,\beta) = \log_2 n$ the graph is walk-regular.

Let us now consider the $i$th entry of the principal eigenvector of the adjacency matrix $\varphi_1(i)$. This index was introduced by Bonacich as the *eigenvector centrality* node $i$ in a social network [22, 23] and it has found important applications in the study of node centrality in complex networks. The following result allows a relationship between the eigenvector centrality and the walk entropy of graphs. When $\beta \to \infty$ we have that $(e^{\beta A})_{ii} \to \varphi_1^2(i) e^{\beta \lambda_1}$ and $Z \to e^{\beta \lambda_1}$, such that $\lim_{\beta \to \infty} p_i(\beta) = \varphi_1^2(i)$. This means that the square of the eigenvector centrality represents the probability of selecting at random a closed walk starting (and ending) at node $i$ at the zero temperature limit of the graph, i.e., when $\beta \to \infty$.

Consequently, the walk entropy for the zero temperature limit is given in terms of the eigenvector centrality

$$S^V(G, \beta \to \infty) = -\sum_i \varphi_1^2(i) \log_2 \varphi_1^2(i). \tag{6}$$

From the properties of the principal eigenvector of the adjacency matrix it is straightforward to realize that $\max_n S^V(G_n, b \to \infty) = \log_2 n$ is attained for any regular graph. On the other hand, when the temperature tends to infinite the walk entropy of any graph attains its maximum. That is, $S^V(G, \beta \to 0) = \log_2 n$.

A graph transformation which has received great attention in the mathematical physics literature is the tensor product of graphs [24-27]. The *tensor product* (also known as the Kronecker product) $G \otimes H$ *of graphs* $G$ and $H$ is the graph whose adjacency matrix is $A(G \otimes H) = A(G) \otimes A(H)$ [28]. Godsil and McKay [14] have proved the following result about the tensor product of two walk-regular graphs.

**Theorem 1** [14]. Let $G$ and $H$ be walk-regular graphs. Then, $G \otimes H$ is walk-regular.

Using this result we prove an additive property for the walk entropy of the tensor product of two graphs.

**Proposition 1.** Let $G$ and $H$ be walk-regular graphs. Then,

$$S^V(G \otimes H) = S^V(G) + S^V(H). \tag{7}$$

*Proof*: Because $G$ and $H$ are walk-regular graphs, $G \otimes H$ is also walk-regular. Then,

$$S^V(G \otimes H) = \log_2 n(G \otimes H) = \log_2[n(G) \cdot n(H)] = \log_2 n(G) + \log_2 n(H) = S^V(G) + S^V(H). \square$$

## 3. Walk entropy of line graphs

The line graph of a graph has been used in several areas of mathematical physics ranging from the Hubbard model to finding partitions in complex networks [29-33]. The line graph $L(G)$ of the graph $G$ is the graph obtained using the following graph transformation. A node of $L(G)$ represents an edge of $G$, and two nodes in $L(G)$ are connected if and only if the corresponding two edges in $G$ shares a common node, i.e., they are adjacent. In short, the line graph $L(G)$ represents the adjacency relationships between the edges in the original graph $G$. For instance, let the nodes $i$ and $j$ be adjacent in $G$, that is, they form an edge $v = \{i, j\}$ in $G$. Then, $v$ is represented as a node in $L(G)$ and its degree $\tilde{k}_v$ is related to the degrees of the nodes $i$ and $j$ in $G$ by $\tilde{k}_v = k_p + k_q - 2$. Some well-known examples of line graphs are the kagome and checkerboard lattices, which are the line graphs of the honeycomb and square lattices, respectively.

The walk-entropy of the line graph $L(G)$ of the graph $G$ is defined by

$$S^V(L(G),\beta) \stackrel{def}{=} -\sum_i \frac{\left(e^{\beta \tilde{A}}\right)_{ii}}{\tilde{Z}} \log_2 \frac{\left(e^{\beta \tilde{A}}\right)_{ii}}{\tilde{Z}}, \tag{8}$$

where $\tilde{A}$ is the adjacency matrix of $L(G)$ and $\tilde{Z} = tr(\exp(\tilde{A}))$. Because $L(G)$ represents the adjacency between the edges in $G$ we can express the walk entropy of the line graph in terms of the adjacency matrix of the graph by using the following expression:

$$S^V(L(G),\beta) = -\sum_{(i,j)\in E} \frac{\left(e^{\beta A}\right)_{ij}}{\sum_{ij}\left(e^{\beta A}\right)_{ij}} \log_2 \frac{\left(e^{\beta A}\right)_{ij}}{\sum_{ij}\left(e^{\beta A}\right)_{ij}}, \tag{9}$$

where $(i,j) \in E$ indicates that the nodes $i$ and $j$ are adjacent in $G$.

Our first result in this section is to show that the probability of finding a walk located at a node of the line graph is related to the average energy of the whole graph. Let the probability that a walk selected at random from the set of all walks in $L(G)$ is located at the node $v$ of $L(G)$ be denoted by $\tilde{p}_v$. Then, $\tilde{p}_v = p_{ij}$ and we have the following result.

**Proposition 2.** The probability $\tilde{p}_v(\beta) = p_{ij}(\beta)$ is related to the average energy $\langle E \rangle$ of the graph by

$$\tilde{p}_v(\beta) = p_{ij}(\beta) = -\frac{2}{\langle E \rangle} \frac{\left(e^{\beta A}\right)_{ij}}{Z}. \tag{10}$$

***Proof***: Let $\tilde{p}_v(\beta) = p_{ij}(\beta)$ be written as

$$\tilde{p}_v(\beta) = p_{ij}(\beta) = \frac{\left(e^{\beta A}\right)_{ij}}{\sum_{(ij)\in E}\left(e^{\beta A}\right)_{ij}} = \frac{2\left(e^{\beta A}\right)_{ij}}{tr(Ae^{\beta A})}. \tag{11}$$

The average energy $\langle E \rangle$ of the graph is defined by

$$\langle E \rangle \stackrel{def}{=} -\frac{1}{Z}\frac{\partial Z}{\partial \beta}. \tag{12}$$

Because $\frac{\partial}{\partial \beta} tr(e^{\beta A}) = tr(Ae^{\beta A})$, we have that

$$\tilde{p}_v(\beta) = p_{ij}(\beta) = 2(e^{\beta A})_{ij} \left[\frac{\partial tr(e^{\beta A})}{\partial \beta}\right]^{-1} = 2(e^{\beta A})_{ij} \left(\frac{\partial Z}{\partial \beta}\right)^{-1} = -\frac{2}{Z} \frac{(e^{\beta A})_{ij}}{\langle E \rangle}. \quad \square \tag{13}$$

In the following results we describe some properties of the walk entropy of the line graph. The first result is trivial and account for the maximum walk entropy in the line graph. The maximum entropy of the line graph $L(G)$ of $G$ is attained when $L(G)$ is walk-regular. The entropy is then $\max_m S^V(L(G)) = \log_2 m$, where $m$ is the number of edges in $G$. Another trivial result is that

$$S^V(L(G), \beta \to \infty) = -\sum_{(i,j) \in E} \frac{\varphi_1(i)\varphi_1(j)}{\ell} \log_2 \frac{\varphi_1(i)\varphi_1(j)}{\ell}, \tag{14}$$

where $\ell = \langle \varphi_1 | A | \varphi_1 \rangle$ is sometime called the graph Lagrangian. It attains the maximum $\max_m S^V(L(G_{n,m}), \beta \to \infty) = \log m$ when the line graph is regular.

We now prove a result analogous to that of Godsil and McKay [14] (Theorem 3) for the case of walk-regular line graphs. First, we need the following auxiliary result.

**Proposition 3.** Let $A \otimes B$ and $C \otimes D$ be two matrices of the same size. Then

$$(A \otimes B) \circ (C \otimes D) = (A \circ B) \otimes (C \circ D),$$

where $\circ$ represents the Schur (also known as Hadamard or entrywise) product.

*Proof*: Let $A = [a_{is}]$ and $C = [c_{sj}]$. By the definition of the Kronecker product, $A \otimes B = [a_{is}B]$ and $C \otimes D = [c_{sj}D]$. Then, the $(i,j)$th block of $(A \otimes B) \circ (C \otimes D)$ is $(a_{is}B) \circ (c_{sj}D) = (a_{is}c_{sj}) \circ (BD)$. Also the $(i,j)$th block of $A \circ C$ is $a_{is}c_{sj}$, which implies that $(A \otimes B) \circ (C \otimes D) = (A \circ B) \otimes (C \circ D)$ as required. $\square$

**Theorem 2.** If $L(G)$ and $L(H)$ are walk-regular line graphs, then the line graph of the tensor product of $L(G)$ and $L(H)$, $L(L(G) \otimes L(H))$ is also walk-regular.

*Proof*: Let $A_G$ designates the adjacency matrix of the graph $G$. Then, $L(G)$ is walk-regular if and only if

$$A_G \circ A_G^k = \alpha A_G, \tag{15}$$

for all $k$, where $\alpha$ is a certain integer number. Then, for $G \otimes H$ we have

$$\left(A_G \otimes A_H\right) \circ \left(A_G \otimes A_H\right)^k = \left(A_G \otimes A_H\right) \circ \left(A_G^{\ k} \otimes A_H^{\ k}\right)$$
$$= \left(A_G \circ A_G^{\ k}\right) \otimes \left(A_H \circ A_H^{\ k}\right) \tag{16}$$

Because $L(G)$ and $L(H)$ are walk-regular line graphs we have

$$\left(A_G \otimes A_H\right) \circ \left(A_G \otimes A_H\right)^k = \left(\alpha A_G^{\ k}\right) \otimes \left(\gamma A_H^{\ k}\right)$$
$$= \alpha\gamma\left(A_G^{\ k} \otimes A_H^{\ k}\right) \quad \square \tag{17}$$
$$= \alpha\gamma\left(A_G \otimes A_H\right)^k.$$

Consequently, we have the following result concerning the walk entropy of the tensor product of two walk-regular line graphs.

**Proposition 4.** Let $L(G)$ and $L(H)$ be walk-regular line graphs. Then,

$$S^V(L(G) \otimes L(H)) = S^V(L(G)) + S^V(L(H)) + 1. \tag{18}$$

***Proof***: Because $L(G)$ and $L(H)$ are walk-regular line graphs, $L(L(G) \otimes L(H))$ is also walk-regular. Then,

$$S^V(L(G) \otimes L(H)) = \log_2 m(L(G) \otimes L(H)) = \log_2[2 \cdot m(G) \cdot m(H)]$$
$$= \log_2 2 + \log_2 m(G) + \log_2 m(H) \quad \square$$
$$= S^V(L(G)) + S^V(L(H)) + 1.$$

## 4. Extremal graphs for the walk entropy

In this section we calculate the walk entropy for all connected graphs with at most 8 nodes (~12,000 graphs) and find those graphs with the minimum walk entropy for a given number of nodes. For instance, for $n = 8$ the graphs with the minimum walk entropy consist of a clique of 5 nodes and three pendant nodes attached at different positions of the clique (see Figure 1). Similar results are obtained for $n = 7$, where the graphs minimizing $S^V(G)$ have a 5-nodes clique and two pendant nodes.

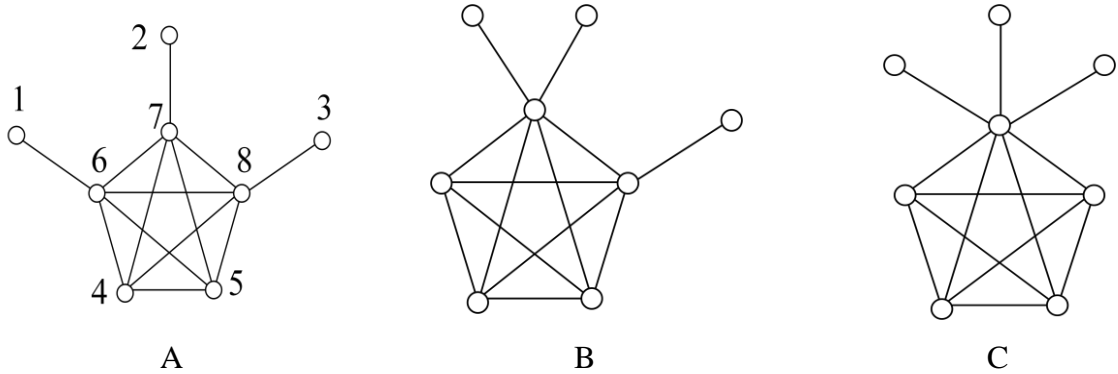

A B C

**Figure 1.** The graphs with $n = 8$ that have the minimum walk entropies.

The main characteristic of these graphs is that they consist of two types of nodes with very different probabilities of finding a random walker at them. The nodes in the clique display very high values of $G_{pp}$, i.e. high probability of finding the random walker on them, while the pendant nodes display very low probabilities of finding the walker. In fact, the probability of finding the walker on the clique is 10 times bigger than that for the nodes in the the pendant ones. In other words, the walker is very much localized on those nodes of the clique (see Figure 2). As a consequence of this 'localization' of the walker, the walk entropy is very small because it is relatively easy to find the random walker in a specific region of the graph. In fact, $S^V(G)$ displays a Pearson correlation coefficient of 0.43 with the inverse participation ratio (IPR), which is a widely used criterion for the localization of states. In contrast, the von Neumann entropy of the graph displays only a Pearson correlation coefficient of 0.27 with IPR. We remind the reader that the inverse participation ratio $I_j$ for the $j$ th eigenstate is defined by

$$I_j = \left(\sum_p |\varphi_j(p)|^4\right)^{-1}, \tag{19}$$

where a fully localized state is characterized by $I_j = 1$, values close to unity indicates strong on-site localization of the $j$ th eigenstate, and $I_j \gg 1$ indicates no localization. This index can be averaged over all eigenstates to obtain:

$$\langle IPR \rangle = n^{-1} \sum_{j=1}^{n} I_j. \tag{20}$$

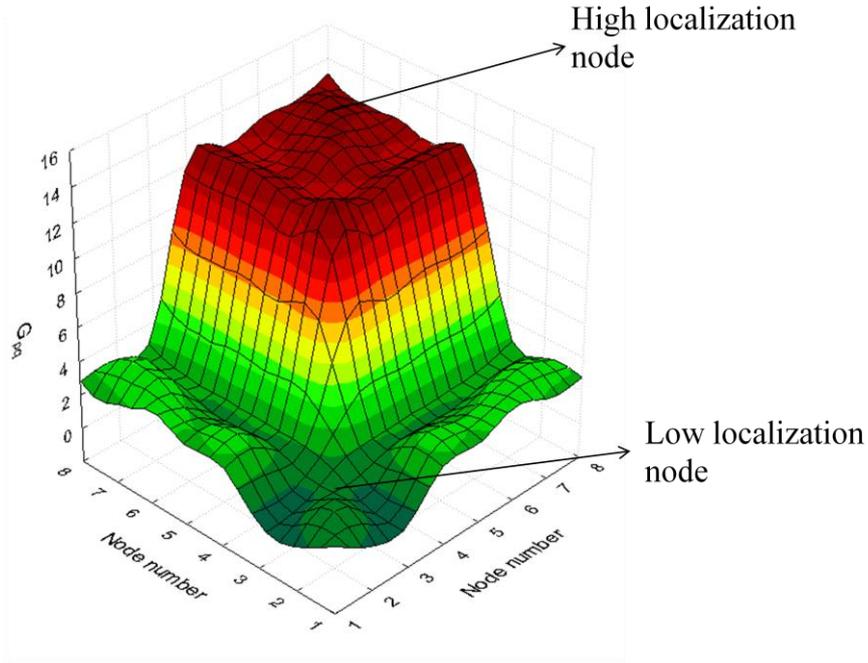

**Figure 2.** Plot of the communicabilities $G_{pq}$ for pairs of nodes in the graph displayed in the Figure 1A.

We now consider the graphs whose line graphs display the minimum walk entropies. In Figure 3 we illustrate these graphs with 8 nodes and 7, 8, 9 and 10 edges, respectively.

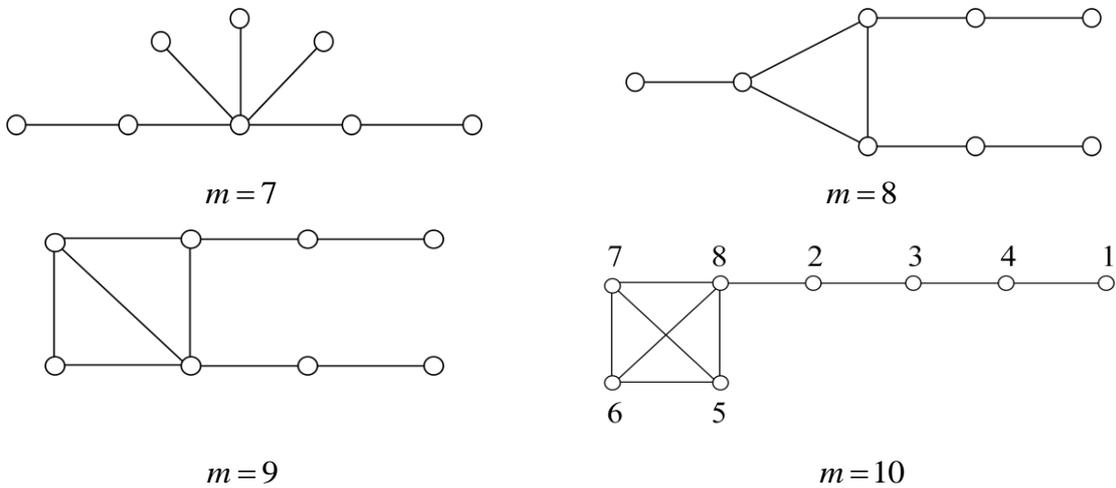

**Figure 3.** The graphs with $n = 8$ and different number of edges whose line graphs have the minimum walk entropies.

When these graphs are transformed into their line graphs, a few nodes (representing edges of $G$) display very large localization of the random walker, while the rest of the nodes in $L(G)$ remain with low localization of the walker. As a consequence, the walk entropy of the line graphs of these graphs is very small. Because the nodes of the line graph represent the edges of the graph we can imagine that those nodes with high localization in $L(G)$ correspond to the hypotentical case in which we trap the random walker just between two nodes in the graph, i.e. in an edge of $G$.

**5. Modulating the walk entropy and localization**

In this section we study how the walk entropies, and accordingly the degree of localization of a random walker, change with the change in the temperature. We concentrate now on the walk entropy of the graphs as the results for the line graphs are quite similar. As we have previously shown $S^V(G, \beta \to 0) = \log n$, which is the maximum that the walk entropy can attain for a graph. Then, by decreasing the temperature from $\beta = 0$ the walk entropy will not increase. However, different behaviours are expected for different kinds of graphs. For the purposes of the current study we can divide the graphs into three different classes. The first class is the class of the walk-regular graphs, here designated as $W^0$. Let $R$ be the class of the regular graphs which are not walk-regular. The rest of the graphs are simply non-regular. Then, let us now analyze the variation of the walk entropy with the inverse temperature for these classes of graphs.

The walk-regular graphs are characterized by the fact that their walk entropy does not change with any change (increase of decrease) of the temperature. In other words, the localization of a random walker in a walk-regular graph is not affected by the changes in the temperature. For these graphs, the probability of finding the walker at a given node is always equal to the inverse of the size of the graph.

Let us now consider the case of the non-regular graphs. These graphs are characterized by the fact that $S^V(G, \beta = 1) > S^V(G, \beta \to \infty)$, which means that they display greater delocalization of the walker at $\beta = 1$ than when $\beta \to \infty$. That is, for a non-regular graph at $\beta = 1$ a decrease of the temperature ($\beta \to \infty$) will increase the localization of the walker at certain nodes of the graph. That is, the walker is 'frozen' at certain nodes of the graph where the probability of finding it is relatively large in comparison with that in the rest of the nodes of the graph. Consequently, the walk entropy of the graph drops to the value given by $-\sum_i \varphi_1^2(i) \log_2 \varphi_1^2(i)$, after which it remains constant (see Figures 4 and 5).

Now, we analyze the behavior of regular graphs which are not walk-regular, $G \in R, G \notin W^0$. Regular graphs are characterized by the fact that $S^V(\beta \to 0) = S^V(\beta \to \infty) = -\sum_i \varphi_1^2(i) \log_2 \varphi_1^2(i)$ and $S^V(\beta = 1) < S^V(\beta \to 0) = S^V(\beta \to \infty)$. This implies that if we increase $\beta$ from $\beta = 0$ the walk entropy decreases as in the case of non-regular graphs, but after a certain point at which the entropy is minimum, it will increase again up to reaching its maximum when $\beta \to \infty$. That is, at $\beta = 0$ ($T \to \infty$) the graph consists of isolated nodes, and the probability of finding the particle at each site is independent of the vertex. As the temperature decreases some localization starts to appear in the regular graph due to its lack of walk regularity. As a consequence, the walk entropy of the graph decreases. Eventually, at a given $\beta$, which depends on the structure of the regular graph, the entropy reaches a minimum. At this point the regular graph displays the highest difference in the localization of the random walker in the nodes of a regular graph. After this point an increase in $\beta$ makes that the first eigenmode (eigenvector associated with the maximum eigenvalue) has a dominant contribution to the probability of finding the walker at a given node. Because the graph is regular all the entries of this eigenvector are the same, which means that the entropy starts to increase again up to reaching its maximum at $-\sum_i \varphi_1^2(i) \log_2 \varphi_1^2(i) = \log n$ (see Figures 4 and 5).

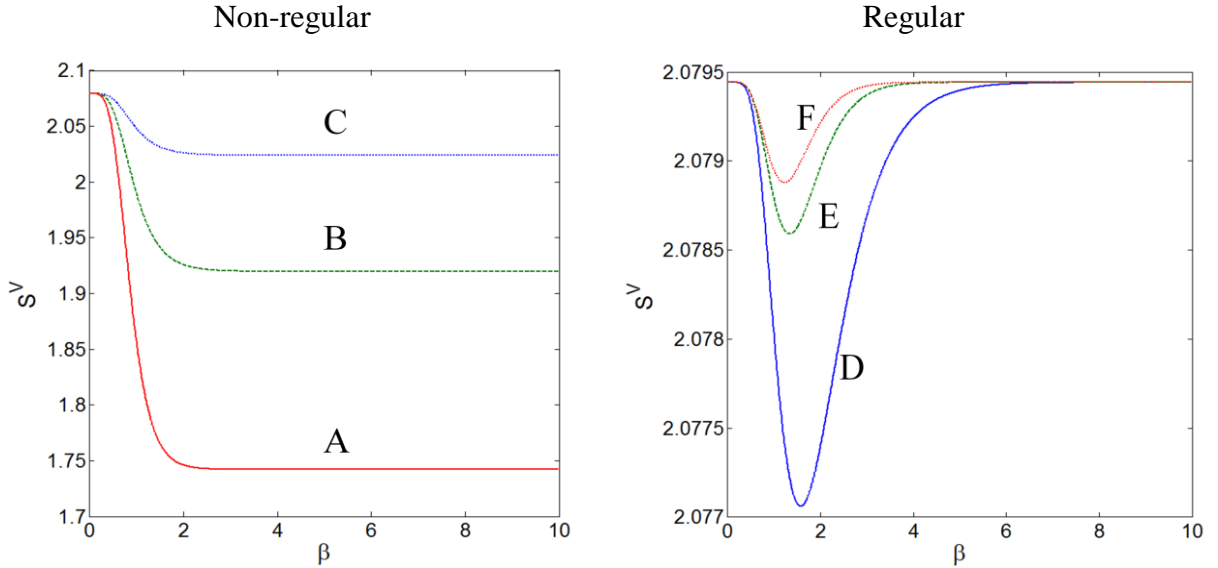

**Figure 4.** Change of the walk entropy with the increase of the inverse temperature for (left) three non-regular graphs and (right) three cubic graphs (regular graphs of degree 3). The structures of the six graphs are illustrated in Figure 5. The plots of the non-regular graphs are: continuous red line (graph A), broken green line (graph B) and dotted blue line (graph C). For the regular graphs the plots are: continuous blue line (graph D), broken green line (graph E), dotted red line (graph F).

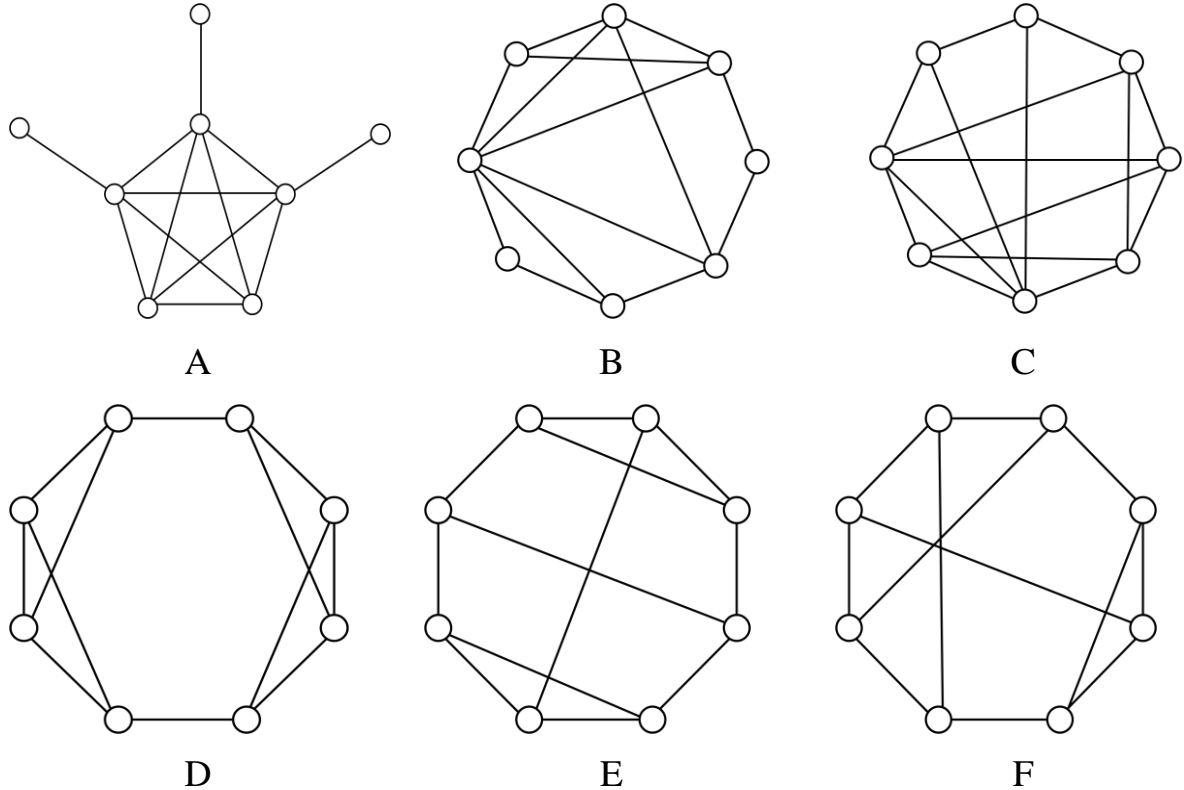

**Figure 5.** Illustration of the graphs for which the change of the walk entropy versus the inverse temperature are plotted in Figure 4.

## 6. Summary

In the present article, we defined the walk entropy for graphs and their line graphs. The walk entropies are defined on the basis of the summation over diagonal and off-diagonal elements of the exponentiated adjacency matrix. One of their notable properties is that it does not depend on the graph size.

The walk entropies indicate how much localized a random walker is on the nodes or edges of a graph, respectively. The maximum delocalization and consequently maximum walk entropy, is attained for walk-regular graphs/line-graphs.

We prove some relations between the walk entropy of graphs/line graphs and those of their tensor products. We also showed the temperature dependence of the walk entropies. Though the walk entropy for walk-regular graphs does not depend on the temperature, it does for other graphs. Particularly interesting is the behavior of the walk entropy of regular, but non-walk-regular graphs, for which the entropy is non-monotonic, taking its maximum at zero temperature as well as at infinite temperature and a minimum at a temperature in between. This means that the walker is delocalized over the graph both at zero and infinite temperature, while it is relatively localized at moderate temperatures.

**Acknowledgements**

EE thanks both CIMAT (Guanajuato) and the Institute of Industrial Science, University of Tokyo for warm hospitality in the periods between February and March, 2013. EE and NH thanks partial financial support from Aihara Innovative Mathematical Modeling Project, the Japan Society for the Promotion of Science (JSPS) through the "Funding Program for World-Leading Innovative R&D on Science and Technology (FIRST Program)," initiated by the Council for Science and Technology Policy (CSTP).

**References**

[1] Estrada E 2011 *The Structure of Complex Networks. Theory and Applications* (Oxford: Oxford University Press)

[2] da F Costa L Rodrigues F A Travieso G and Villas Boas P R 2007 *Adv. Phys.* **56** 167-242

[3] Bose S 2003 *Phys. Rev. Lett.* **91** 207901

[4] Marzuoli A and Rasetti M 2005 *Ann. Phys.* **318** 345-407

[5] Kay A 2011 *Phys. Rev. A* **84** 022337

[6] Chiribella G D'Ariano G M and Perinotti P 2009 *Phys. Rev. A* **80** 022339

[7] Godsil C Kirkland S Severini S and Smith J 2012 *Phys. Rev. Lett.* **109** 50502

[8] Estrada E 2013 arXiv:1302.4378

[9] Dehmer M and Mowshowitz A A 2011 *Inform. Sci.* **181** 57-78

[10] Passerini F and Severini S 2009 *Int. J. Agent Tech. Syst.* **1** 58-67

[11] Estrada E and Hatano N 2007 *Chem. Phys. Lett.* **439** 247-251

[12] Bianconi G 2009 *Phys. Rev. E* **79** 036114

[13] Estrada E Hatano N and Benzi M 2012 *Phys. Rep.* **514** 89-119

[14] Godsil C D and McKay B D 1980 *Lin. Alg. Appl.* **30** 51-61

[15] Gamble J K Friese M Zhou D Joynt R and Coppersmith S N 2010 *Phy. Rev. A* **81** 052313

[16] Rudinger K Gamble J K Bach E. Friesen M Joynt R and Coppersmith S N 2012 arXiv:1207.4535v1

[17] Rudinger K Gamble J K Wellons M Bach E Friesen M Joynt R and Coppersmith S N 2012 arXiv:1206.2999v2

[18] Travaglione B C and Milburn G J 2002 *Phys. Rev. A* **65** 032310

[19] Bernasconi A Godsil C and Severini S 2008 arXiv:0808.0510

[20] Hashimoto Y Hora A and Obata N 2003 *J. Math. Phys.* **44** 71

[21] Estrada E and Hatano N 2008 *Phys. Rev. E* **77** 036111

[22] Bonacich P 1972 *J. Math. Sociol.* **2** 113-120

[23] Bonacich P 1987 *Am. J. Soc.* **92** 1170-1182

[24] Braunstein S L Ghosh S and Severini S 2006 *Ann. Comb.* **10** 291-317

[25] Golénia S and Schumacher Ch 2010 arXiv:1005.0165

[26] Klavžar S and Severini S 2010 *J. Phys. A: Math. Theor.* **43** 212001

[27] Wang Z and Wang Z 2007 J. Comb. **14** R40

[28]     Imrich W and Klavžar S 2000 *Product graphs. Structure and recognition* (New York: Wiley)

[29]     Mielke A 1991 *J. Phys. A: Math. Theor.* **24** 3311

[30]     Lu L 2009 *J. Phys. A: Math. Theor.* **42** 265002

[31]     Motruk J and Mielke A 2012 *J. Phys. A: Math. Theor.* **45** 225206

[32]     Pakoński P Tanner G and Życzkowski K 2003 *J. Stat. Phys.* **111** 1331

[33]     Evans T S and Lambiotte R 2009 *Phys Rev E* **80** 016105